\documentclass[11pt,twoside]{article}

\usepackage{asp2006}

\usepackage{graphicx}
\usepackage[breaklinks=true]{hyperref}
\bibpunct[,]{(}{)}{;}{a}{}{,} 

\begin{document}
\title{Stellar Collisions in Young Clusters: \\Formation of (Very) Massive Stars?}   
\author{Marc Freitag}   
\affil{Institute of Astronomy, Madingley road, Cambridge CB3\ 0HA, UK}    

\begin{abstract} 
In young star clusters, the density can be high enough and the
velocity dispersion low enough for stars to collide and merge with a
significant probability. This has been suggested as a possible way to
build up the high-mass portion of the stellar mass function and as a mechanism
leading to the formation of one or two very massive stars ($M_\ast >
150\,{\rm M}_\odot$) through a collisional runaway.  I quickly review
the standard theory of stellar collisions, covering both the stellar
dynamics of dense clusters and the hydrodynamics of encounters between
stars. The conditions for collisions to take place at a significant
rate are relatively well understood for idealised spherical cluster
models without initial mass segregation, devoid of gas and composed of
main-sequence (MS) stars. In this simplified situation, 2-body
relaxation drives core collapse through mass segregation and a
collisional phase ensues if the core collapse time is shorter than the
MS lifetime of the most massive stars initially present. The outcome
of this phase is still highly uncertain. A more realistic situation is
that of a cluster still containing large amounts of interstellar gas
from which stars are accreting. As stellar masses increase, the
central regions of the cluster contracts. This little-explored
mechanism can potentially lead to very high stellar densities but it
is likely that, except for very rich systems, the contraction is
halted by few-body interactions before collisions set in. 
A complete picture, combining both scenarios, will need to address many uncertainties, including the
role of cluster sub-structure, the dynamical effect of
interstellar gas, non-MS stars and the structure and evolution of
merged stars.
\end{abstract}



\section{Collision and Interaction Timescales}


We first consider the timescale for direct collisions between
stars. We assume two populations of stars, with stellar masses
$m_1$ and $m_2$, stellar radii $r_1$ and $r_2$, and Maxwellian
velocity distributions of (1D) dispersions $\sigma_1$ and $\sigma_2$.
The mean time for a star of type 1 to collide with a star of
type 2 is
\begin{equation}
t_{\rm coll} = \left\{\sqrt{8\pi}n_2\sigma_{\rm rel}(r_1+r_2)^2
\left[1+\frac{G(m_1+m_2)}{\sigma_{\rm rel}^2(r_1+r_2)}\right]\right\}^{-1},
\label{eq.tcoll}
\end{equation}
with $\sigma_{\rm rel}^2=\sigma_1^2+\sigma_2^2$ \citep{BT87_MFreitag}. The
second term in the square brackets originates in
the mutual gravitational attraction of the stars. This gravitational
focusing dominates over the purely geometrical cross section
(represented by the term $1$ in brackets) unless the velocity
dispersion is similar or higher than the escape velocity from the
surface of a star, i.e.\ $500-1000\,{\rm km}\,{\rm s}^{-1}$. Such
relative velocities are presumably only reached in the vicinity of
massive black holes \citep[e.g.,][]{FB02b_MFreitag,FB05_MFreitag}. For other
situations, the collision time can be written
\begin{equation}
t_{\rm coll} \simeq 
5\,{\rm Gyr}\,\frac{10^6 {\rm pc}^{-3}}{n_\ast}
\frac{\sigma_{\rm rel}}{10\,{\rm km}\,{\rm s}^{-1}}
\frac{2\,{\rm R}_\odot}{r_1+r_2}
\frac{2\,{\rm M}_\odot}{m_1+m_2}.
\label{eq.tcoll2}
\end{equation}
In Fig.~\ref{fig.DensColl}, I show (with dashed lines) the stellar
densities required for a massive star to experience on average one
collision during the MS. Typical values are in excess
of $10^7\,{\rm pc}^{-3}$ or $10^6\,{\rm pc}^{-3}$ for a star of
$10\,{\rm M}_\odot$ or $120\,{\rm M}_\odot$, respectively.

Stars passing each other within a few stellar radii at low velocities
can dissipate enough orbital energy in tides to form a bound binary
\citep*[e.g.,][]{FPR75_MFreitag,KL99_MFreitag}. Such a ``tidal binary'', being born very
tight, is likely to merge quickly, or, at least, to enter a
common-envelope phase \citep{DD06_MFreitag}. The merger can also be induced by
the perturbation of a passing star \citep{BB05b_MFreitag}. In any case, because
tidal dissipation decreases steeply with increasing closest-approach
distance, formation and merging of tidal binary can reduce the
effective value of the collision time by a factor $\sim 3-5$ at
most.


A more promising way to get interesting collision rates is to
consider binary interactions. The timescale over which a binary with
separation $a$ and mass $m_{\rm bin}$ suffers from a close interaction
with a field star is simply
\begin{equation}
t_{\rm bin,inter} \simeq 
30\,{\rm Myr}\,\frac{10^6 {\rm pc}^{-3}}{n_\ast}
\frac{\sigma_{\rm rel}}{10\,{\rm km}\,{\rm s}^{-1}}
\frac{1\,{\rm AU}}{a}
\frac{3\,{\rm M}_\odot}{m_{\rm bin}+m},
\label{eq.tbininter}
\end{equation}
where $m$ is the individual mass of the field stars and $n_\ast$ their
number density. For a sufficiently hard binary, i.e.\ when the total
orbital energy of the binary and impactor system is negative, most
close interactions are ``resonant''. This means that the three stars
orbit each other in a chaotic fashion for several dynamical times
within a volume of order $a^3$
until this meta-stable system decays into a binary and single star if
stars were point masses\footnote{Binary-binary interactions occur in a
similar fashion.}
\citep[e.g.,][]{HH03_MFreitag}. There is in fact a high probability that two of the stars 
collide and merge during a resonant interaction. For instance,
\citet{FregeauEtAl04_MFreitag} showed that, in the hard regime, the cross section 
for a collision to occur during an encounter of a circular, $a=1\,$AU
binary consisting of two $1\,{\rm M}_\odot$ MS stars with a similar
star is about $\frac{3}{2}\pi a G{\rm M}_\odot v_{\rm rel}^{-2}$,
corresponding to a hard sphere of radius $a/4$. If massive binaries
are born with eccentricities $\approx 0.9$, as a result of accretion
of interstellar gas, more distant, non-resonant, interactions with
passing stars can suffice to cause a merger, as suggested by
\citet{BB05b_MFreitag}.

At relative velocities below about $5\,{\rm km}\,{\rm s}^{-1}$,
encounters involving two massive stars, one or both of them
surrounded by a massive disc, can result into a bound binary if the
closest approach distance $d$ is smaller than a few tenth of the disc
radius, which corresponds typically to $d\approx 10-100\,$AU
\citep{MB07a_MFreitag}. With $d$ substituted for $a$, Eq.~\ref{eq.tbininter} 
suggests that this mechanism might be responsible for
the high proportion of massive stars with a companion of similar mass
in clusters such as the ONC, despite the short life time of
massive discs \citep{MB07b_MFreitag,PO07b_MFreitag}.



\section{The Outcome of Stellar Collisions}


Collisions between single MS stars have been the object of extensive
numerical studies, starting with the 2--D grid simulations of\
\citet{SC72_MFreitag}, with most works based on the Smoothed Particle
Hydrodynamics (SPH) method
(e.g., \citealt{FB05_MFreitag}; see the
contributions in \citealt{Shara02} and the ``MODEST'' web
pages\footnote{``Modelling Dense Stellar
Systems''; see \url{http://www.manybody.org/modest/}. Follow the link
``WG4'' for the pages on collisions.} for more references on stellar
collisions).  At low velocities, typical of globular clusters or young
clusters, any collision between MS stars lead to a merger with very
little mass loss ($<10\,$\%, typically a few \%). Neglecting the
possibility of the formation of tidal binaries, stars can therefore be
treated as sticky spheres to good approximation. 


In this velocity regime, one can determine how the mass elements of
the parent stars sort themselves in the collision product, using a
algorithm based on Archimedes' principle, to determine the internal
structure of the merger product with good accuracy, avoiding the
computational cost of an hydrodynamical simulation
\citep{LWRSW02_MFreitag,GLPZ07_MFreitag}. In 
case of collisions between mass of unequal masses, most of the
material from the lower object sinks to the centre of the merged
star. 

One major remaining uncertainty is the evolution of collision products
\citep{SLBDRS97_MFreitag,SFLRW00_MFreitag,SADB02_MFreitag,GP07_MFreitag}. When the merger has settled down to a new 
hydrodynamical equilibrium, it is significantly swollen compared to
the size of a MS star of the same mass, and typically spinning at a
rate close to breakup. Some efficient mechanism, such as winds or
magnetic locking to a disc, has to operate to shed most of this
angular momentum but comparatively little mass as the star contracts
back to the MS, lest it be nearly completely ground down by
``centrifugal evaporation'' \citep*{SAD05_MFreitag}.


From consideration on cross section and the fact that most massive MS
stars are members of binaries (or higher-order systems), it seems that
most collisions in a young cluster occur during binary
interactions. Unfortunately, because of the added numerical complexity
and the enormous parameter space to be explored, only very few
researchers have carried out hydrodynamical simulations of such
interactions \citep{GH91_MFreitag,LombardiEtAl03_MFreitag}. These works made it clear
that {\em multiple} mergers are relatively likely during a binary
interactions due to the increased size of collision products (see also
\citealt{FregeauEtAl04_MFreitag}).


In a very young cluster, one has to consider collisions involving
stars still on the pre-MS. Low-mass stars ($m<5-10\,{\rm M}_\odot$)
take much more time than massive ones to contract to the MS and can
therefore be significantly larger that massive stars
\citep[e.g.,][]{Palla02}.
Only few SPH simulations involving pre-MS stars have been published so
far \citep{ZB02_MFreitag,LS05_MFreitag,DBBBT06_MFreitag}. They show how, in a close encounter with a
massive star, a low-mass pre-MS star is tidally disrupted and end up
forming a disc around the massive star. As mentioned above, such
massive discs offer a very large cross section for the formation of
tight binaries which may later merge, a sequence of events
investigated by \citet{DBBBT06_MFreitag}.

\section{First Scenario: Relaxation-Driven Core Collapse}
\label{sec.relax}

\begin{figure}
\centering
\resizebox{0.99\hsize}{!}{
\includegraphics[bb=19 146 592 716,clip]{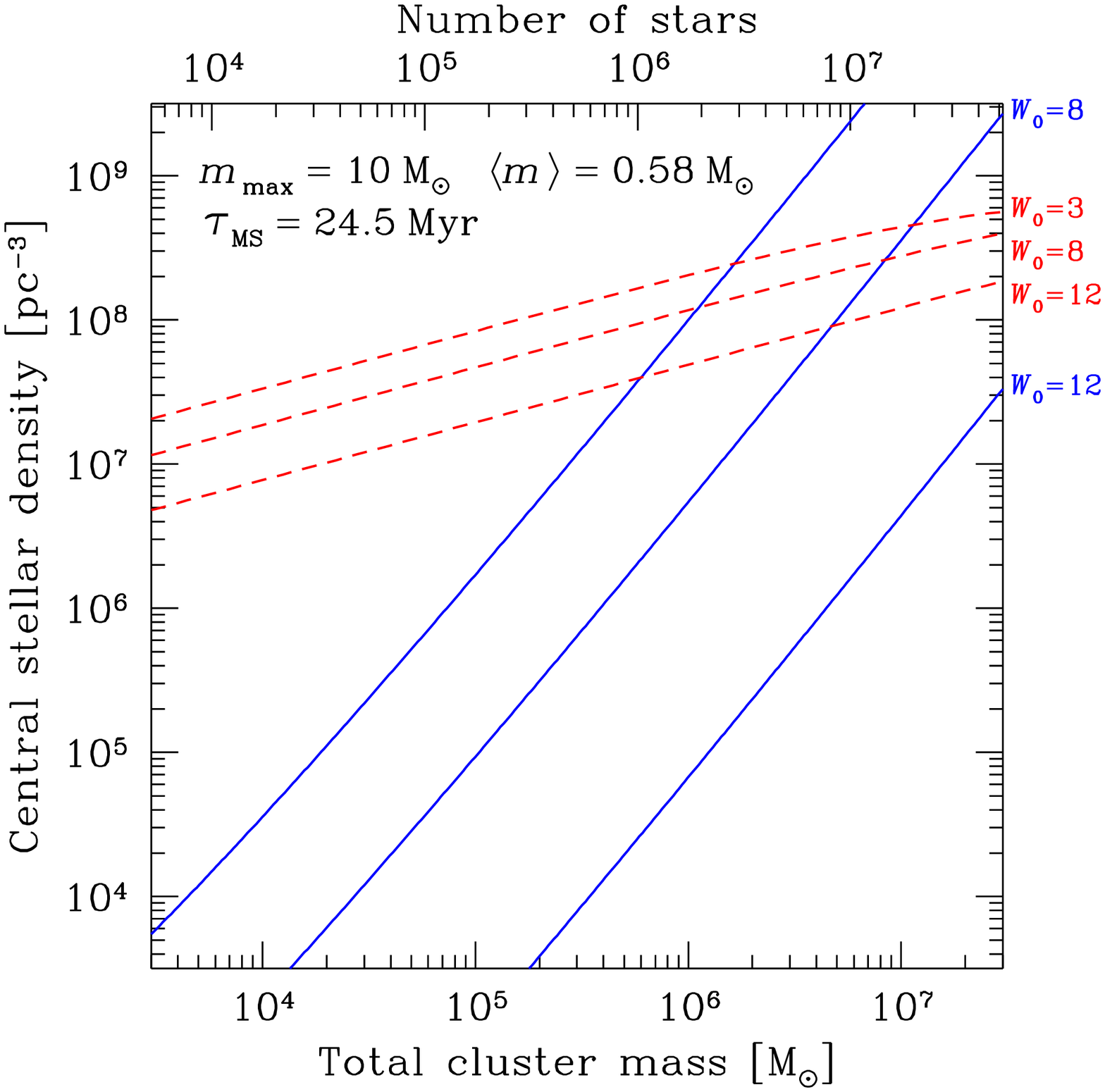}
\includegraphics[bb=86 146 589 716,clip]{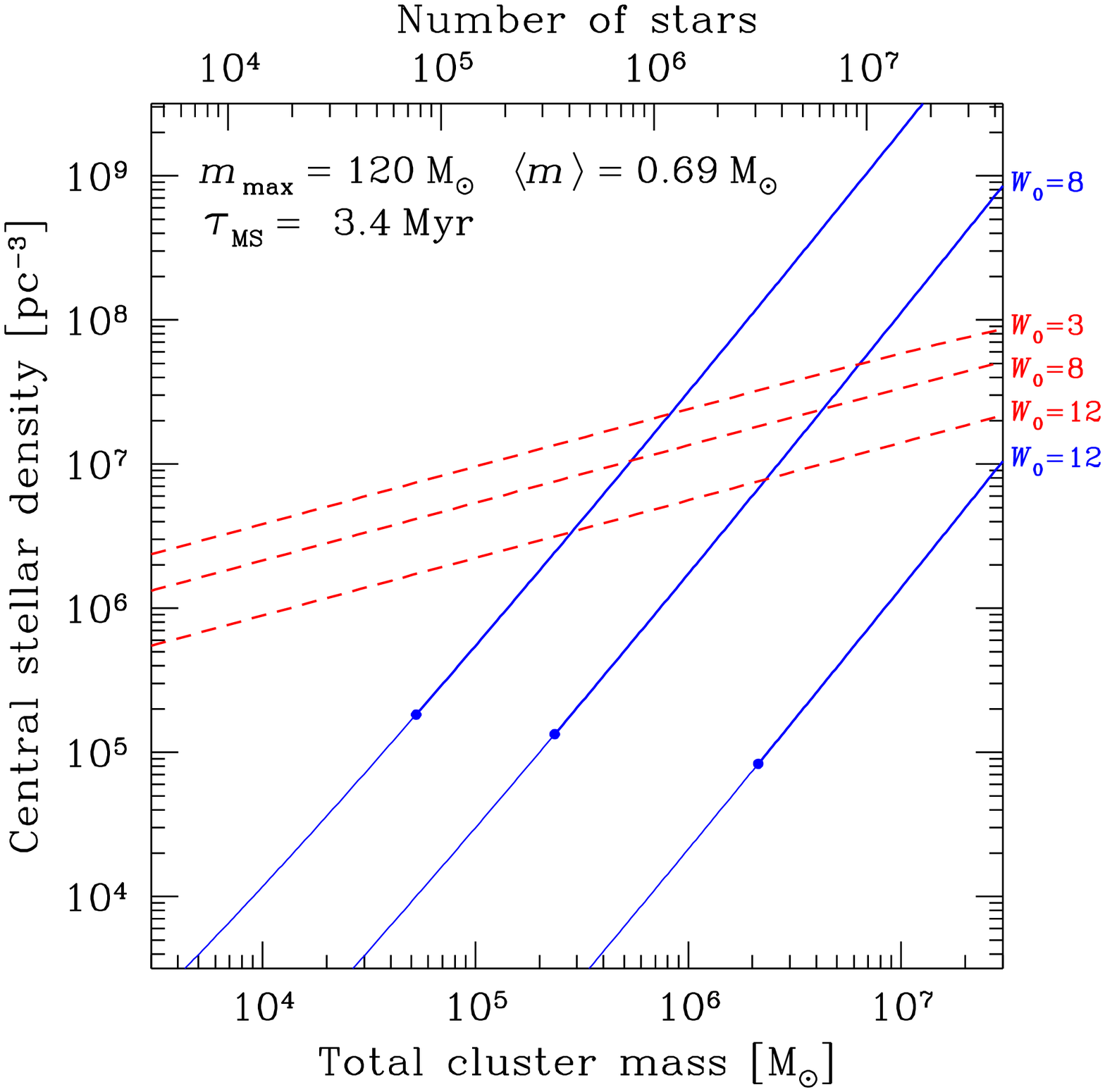}}
\caption[]{Stellar densities required for high collision rate. The dashed lines 
show the density at which the collision time $t_{\rm coll}$ for a star
of mass $m_{\rm max}=10\,{\rm M}_\odot$ (left panel) or $120\,{\rm
M}_\odot$ (right panel) is equal to its MS lifetime $\tau_{\rm MS}$,
at the centre of a non-evolving King-model cluster with $W_0=3,8,12$
(from top to bottom). The dependence on cluster mass (and on $W_0$)
comes from the dependence of $t_{\rm coll}$ on the velocity
dispersion. The solid lines show the central density corresponding to
a core-collapse time $t_{\rm cc}$ equal to $\tau_{\rm MS}$. To
estimate $t_{\rm cc}$, we assume a Salpeter mass function 
from $0.2\,{\rm M}_\odot$ to $m_{\rm max}$. 
The dots on the solid
lines (right panel) are a rough estimate of the number of stars 
above which our estimate of $t_{\rm cc}$ is robust.
\label{fig.DensColl}}
\end{figure}

\begin{figure}
\centering
\resizebox{0.6\hsize}{!}{
\includegraphics{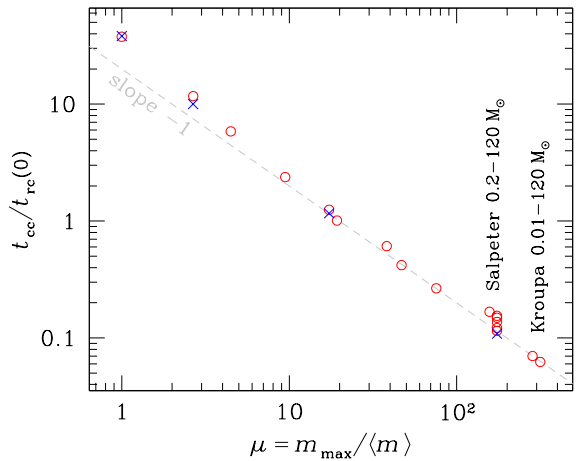}}
\caption[]{Core collapse times for star clusters with a variety of mass functions. 
Shown are previously unpublished results of Monte-Carlo simulations by
the author (circles) and of $N$-body simulations by H.~Baumgardt
(crosses). In most cases, a Plummer model with a MF $dN/dm\propto
m^{-2.35}$ was used. 
\label{fig.tcc}}
\end{figure}



Several authors have studied how 2-body relaxation can drive the
evolution of a cluster to a stage of very high central density,
leading to stellar collisions
(\citealt{PZMMMH99_MFreitag,PZMcM02_MFreitag,PZBHMM04_MFreitag}; \citealt*{GFR04_MFreitag,FreitagRB06_MFreitag,FreitagGR06_MFreitag,GFR06_MFreitag,GGPZ07_MFreitag}).
Two-body relaxation is the process through which stars in a populous
cluster ($N\gg 10$) exchange energy and angular momentum over a time
scale significantly longer than their orbital time, by deflecting each
other's trajectory\footnote{In a smaller group, 2-body relaxation
cannot be defined in a useful way because the energy and angular momentum
of a star are generally not conserved on a dynamical time.}. Roughly
speaking, the relaxation time is the timescale for orbital parameters
of a an ``average star'' to change completely as a result of a
very large number of uncorrelated small-angle hyperbolic 2-body
encounters with all other stars \citep[e.g.,][]{BT87_MFreitag}. Its
local value is given by
\begin{equation}
t_{\rm rlx}\simeq\frac{0.339}{\Lambda}\frac{\sigma^3}{G^2\langle m\rangle^2
n_\ast}
\simeq 2\,{\rm Myr}\,\frac{10}{\Lambda}\frac{10^6 {\rm pc}^{-3}}{n_\ast}
\left(\frac{\sigma}{10\,{\rm km}\,{\rm s}^{-1}}\right)^3
\left(\frac{1\,{\rm M}_\odot}{\langle m\rangle}\right)^2,
\end{equation}
where $\Lambda\simeq\ln(0.02 N)$, $N$ is the number of stars in the cluster, $\sigma$ the 1D velocity
dispersion and $\langle m\rangle$ the mean stellar mass.

In a cluster with a mass function (MF), the massive stars lose kinetic energy in favour of
lower-mass objects through 2-body relaxation. This would eventually
lead to energy equipartition if it was not for the cluster's
self-gravity. As they lose energy, the most massive stars concentrate
to the centre, a process known as dynamical mass segregation and which
reduces to dynamical friction in the limit of very large mass
ratio. For any reasonable MF the central subsystem of massive stars
becomes self-gravitating before equipartition is established. This
makes thermal equilibrium impossible to achieve because the massive
subsystem gets denser and hotter at an accelerated rate as it gives
up energy to the rest of the cluster, an instability first predicted
by \citet{Spitzer69}. Using the Monte-Carlo stellar dynamical method
invented by \citet{Henon71a,Henon71b}, \citet{GFR04_MFreitag} showed that in
rich clusters this mass-segregation-induced core collapse occurs
on a timescale $t_{\rm cc}$ dictated by the initial {\em central}
relaxation time $t_{\rm rc}(0)$, largely independently of the cluster
structure,
\begin{equation}
t_{\rm cc} \approx 2\,t_{\rm rc}(0)\frac{10}{\mu}\mbox{\ \ for\ \ }
\mu\equiv\frac{m_{\rm max}}{\langle m\rangle} > 5.
\end{equation}
This relation is shown in Fig.~\ref{fig.tcc} for a set of cluster
simulations.

The solid lines in Fig.~\ref{fig.DensColl} show what is the minimum
initial central density of a cluster of given mass and concentration
for the core collapse time to be shorter than the MS lifetime
$\tau_{\rm MS}$ of the most massive stars. In Monte-Carlo simulations
where this is the case, the core collapse proceeds until the massive stars
in the central region start colliding, either as single stars if there
are no primordial binaries (a rather unrealistic idealisation) or
during binary interactions. When $t_{\rm cc}>\tau_{\rm MS}$, the gas
lost by massive star in the post-MS evolution causes the core of the
cluster to re-expand before any significant number of collisions can
occur.

Simulations carried out with the more accurate but very time-consuming
direct $N$-body algorithm in the regime $\mu\ge 40$ and
$N<\mbox{few}\times10^5$ indicate a more constraining condition for a
collisional phase to occur \citep[e.g.,][]{PZBHMM04_MFreitag}. This is probably
because the type of core collapse described here can only occur if
there are initially a sufficient number of stars with masses close to
$m_{\rm max}$ in a region not much larger than the core. In
Fig.~\ref{fig.DensColl} we indicate with dots on the solid lines the
cluster masses below which there are on average fewer than 5 stars
with a mass between $0.5\,m_{\rm max}$ and $m_{\rm max}$ in the core
initially. This is not a significant limitation for $m_{\rm
max}=10\,{\rm M}_\odot$.

The outcome of the collisional phase is still a subject of
debate. Dynamical simulations which only take into account collisional
mass loss and assume that the merger product returns to the MS
immediately after a collision show a runaway growth of one or two
stars to masses $m_{\rm ra}$ of a few hundreds or thousands ${\rm
M}_\odot$. Such ``very massive stars'' (VMSs) have been suggested as
progenitors of intermediate-mass black holes but it seems that a VMS
with non-negligible metallicity, if left to evolve on its own, will
lose most of its mass by stellar winds and produce only a rather
low-mass remnant \citep{BVBV07_MFreitag,YungelsonEtAl07_MFreitag}. The collisional
growth can also be self-limited if the average time between two
mergers becomes shorter than the thermal timescale of the VMS, as seen
in Monte-Carlo simulations of very rich cluster of single stars
\citep{FreitagGR06_MFreitag}. Then the VMS can not contract back 
to the MS and might instead become too diffuse to stop impactors.

It is in any case unlikely that this scenario can lead to the
formation of a continuous spectrum of massive stars. Simulations as
well as simple mathematical models suggest that one (possibly two) VMS
detaches itself from the initial MF with very few, if any, stars
populating the gap between $m_{\rm max}$ and $m_{\rm ra}$.

\section{Second Scenario: Accretion-Driven Core Collapse}

The relaxation-driven scenario is based on a simplistic
model of young clusters, neglecting the role of non-spherical
substructure, initial mass segregation and interstellar gas. The last
point is of particular importance because during the first Myr or so
of their lives, most of the mass of clusters is in gas rather than
stars. When the residual gas is expelled through ionisation by the new born
massive stars or, for the most massive clusters, by the first SN
explosion, the cluster reacts by expanding considerably, probably to
the point of complete dispersion for most low-mass clusters
\citep*{GB01_MFreitag,KAH01_MFreitag,GB06_MFreitag,BK07_MFreitag}. If a cluster which 
remains bound is a factor $X_R$ larger and a factor $X_M$ less massive
than in the embedded phase, with $X_R\approx 10$ and $X_N\approx 0.3$
being reasonable values, one can estimate that its relaxation time has
{\em increased} by a factor $X_{\rm rlx}\approx X_R^{3/2}X_M^{\nu}$
with $\nu\ge 0.5$ depending on what fraction of the stars are lost.
$X_{\rm rlx}$ could be as large as $\sim 20$, indicating that the
segregation-driven core collapse time could be much shorter in the
embedded phase than estimated from observations of young, gas-free
clusters.  Incidentally, the collision time increases by $X_{\rm
coll}\approx X_R^{5/2}X_M^{\nu'}$ with $\nu'\le 0.5$, a factor which
can reach hundreds.

Such estimates suggest that the focus of the study of collisions in
young clusters should be shifted to the embedded phase and to stars
that are still in the process of formation, rather than MS objects. In
this context, another process can lead to a strong increase of the
stellar density, possibly up to a collisional phase
(\citealt*{BBZ98_MFreitag,BB02_MFreitag}). The orbits of stars accreting from the
interstellar gas shrink as a consequence of conservation of linear
momentum. If the gas has zero net momentum, does not dominate the
potential, and is accreted on a timescale long compared to orbital
time, the size of the stellar system will contract like $R_{\rm sys}
\propto M_{\rm sys}^{-3}$, where $M_{\rm sys}$ is the combined,
increasing mass of the stars. The above assumptions should be valid
for the core of a forming cluster or sub-cluster, rather than the
whole cluster, $N_{\rm sys}<N$. The overall contraction of the system is eventually
halted by small-$N$ stellar dynamical effects, in particular energy
release by hard binaries. The maximum stellar density that can be
reached is estimated to be $\rho_{\rm max}\equiv \langle m\rangle n_{\rm
max}\approx \left(M_{\rm tot}/\langle m\rangle\right)^2
\bar{\rho}$, where $\langle m\rangle$ is the mean stellar 
mass reached at that time in the contracting system, $M_{\rm tot}$ the,
total mass of the parent proto-cluster and $\bar{\rho}$ its average
mass density. This density appears to be high enough for collisions to affect a
significant fraction of stars only in rather rich systems, $N_{\rm
sys} \gg 100$ (Clarke \& Bonnell, in preparation).


Clearly the role of collisions in massive star formation or, more
generally, in the dynamisc of young clusters can not be assessed
independently of all other processes taking place in these complex
systems. In this contribution, I have surveyed a few aspects that
require consideration (see, e.g., \citealt{BZ05_MFreitag,ZY07_MFreitag} for other
discussions of this subject). It seems that a complete picture will
only be reached by means of numerical simulations combining stellar
dynamics, the physics of the interstellar gas and the formation and
early evolution of single and binary stars.

\acknowledgements 
It is a pleasure to thank Houria Belkus, Ian Bonnell, Cathie Clarke,
James Lombardi and Hans Zinnecker for discussions and data used to
prepare the talk on which this paper is based.  My work is founded
through the STFC theory rolling grant to the Institute of Astronomy in
Cambridge.



\end{document}